\documentclass[12pt,preprint]{aastex}

\usepackage{epsfig}
\usepackage{graphicx}
\usepackage{epstopdf}
\usepackage{amsmath}

\usepackage{rotating}

\usepackage{natbib}
\begin{document}

\title{Calibration of the MEarth Photometric System: Optical Magnitudes and Photometric Metallicity Estimates for 1802 Nearby M-dwarfs}
\author{Jason A. Dittmann$^{1}$, Jonathan M. Irwin$^{1}$, David Charbonneau$^{1}$, Elisabeth R. Newton$^{1}$}
\affil{[1] Harvard-Smithsonian Center for Astrophysics, 60 Garden St., Cambridge, MA, 02138} 

\begin{abstract}
The MEarth Project is a photometric survey systematically searching the smallest stars nearest to the Sun for transiting rocky planets. Since 2008, MEarth has taken approximately two million images of 1844 stars suspected to be mid-to-late M dwarfs. We have augmented this survey by taking nightly exposures of photometric standard stars and have utilized this data to photometrically calibrate the $MEarth$ system, identify photometric nights, and obtain an optical magnitude with $1.5\%$ precision for each M dwarf system. Each optical magnitude is an average over many years of data, and therefore should be largely immune to stellar variability and flaring. We combine this with trigonometric distance measurements, spectroscopic metallicity measurements, and 2MASS infrared magnitude measurements in order to derive a color-magnitude-metallicity relation across the mid-to-late M dwarf spectral sequence that can reproduce spectroscopic metallicity determinations to a precision of 0.1 dex. We release optical magnitudes and metallicity estimates for 1567 M dwarfs, many of which did not have an accurate determination of either prior to this work. For an additional 277 stars without a trigonometric parallax, we provide an estimate of the distance assuming solar neighborhood metallicity. We find that the median metallicity for a volume limited sample of stars within 20 parsecs of the Sun is [Fe/H] = $-0.03 \pm 0.008$, and that 29 / 565 of these stars have a metallicity of [Fe/H] = $-0.5$ or lower, similar to the low-metallicity distribution of nearby G-dwarfs. When combined with the results of ongoing and future planet surveys targeting these objects, the metallicity estimates presented here will be important in assessing the significance of any putative planet-metallicity correlation. 

\end{abstract}

\section{Introduction}
M dwarfs are the most common type of star in the universe, making up approximately 70\% of all stars. Despite their ubiquity, they remain poorly understood. M dwarfs span a large range in mass, from 0.6 M$_\sun$ to the hydrogen burning mass limit at approximately 0.08 M$_\sun$. This factor of 7.5 in mass ratio from the early to late M dwarfs is comparable to the mass ratio between the entirety of the AFGK spectral sequence. Over this mass range, the photospheric effective temperature of these stars cools such that molecular species (TiO, VO, CO, FeH, among others) begin to dominate the optical spectrum, significantly depressing the optical flux as well as cluttering the spectrum with many overlapping lines and bands. This effect makes line-by-line spectral analysis in these stars challenging. Modeling these features in bulk is also extremely difficult because accurate and complete line lists for many identified species are unavailable and some lines have yet to be identified. 

In order to get around this inherent difficulty, progress has been made in calibrating empirical relations using M dwarf stars in wide visual binaries with better understood F, G, or K primary stars. Since they are physically associated, they likely formed out of material with the same bulk composition. Therefore the metallicity measured for the FGK primary star can reasonably be assumed to be the metallicity of the M dwarf, and an empirical relation can be calibrated. \citet{B05} conducted a study of 20 M dwarfs binaries with a solar type primary and were able to calibrate a photometric metallicity relation using the M dwarf's $M_K$ magnitude and ($V - K$) color with a precision of 0.2 dex.  A further study by \citet{johnjohn_metallicity} found that this calibration systematically underestimated M dwarf metallicities by nearly 0.3 dex for M dwarfs with high metallicities, which was outside the range in the \citet{B05} calibration sample. However, they confirmed a key finding of \citet{B05}: the optical $-$ infrared color of an M dwarf is significantly affected by the metallicity of the star.  \citet{Schlaufman_Laughlin_12} expanded this work and attempted to correct systematic issues with both the \citet{B05} and the \citet{johnjohn_metallicity} calibrations. They altered the calibration on the edges of the metallicity range investigated, as well as expanded the calibration to a wider range of spectral types and increased the size of the calibration sample. The standard deviation of their fit was also reduced to 0.15 dex. All three of these calibrations were tested by \citet{Neves12} on an independent sample of FGK - M common proper motion binaries, who found that the relation in \citet{Schlaufman_Laughlin_12} performed the best, with an RMS residual of $0.19$ dex. 

Considering spectroscopy, recent progress has been made by moving to near infrared spectra, where the density of lines is much sparser and it is easier to measure a reliable continuum value. \citet{babs} found that the combination of the Na I doublet and the Ca I triplet in the infrared $K$-band are reliable metallicity indicators, provided one accounts for the temperature of the star with a temperature indicator, such as the H$_2$O-K2 index. \citet{babs} calibrated this correlation using common proper motion pairs of M dwarfs paired with main sequence FGK stars. This calibration was then updated by \citet{babs12} and applied to 133 solar neighborhood M dwarfs. A similar calibration in the $H$-band utilizing 22 common proper motion pairs was created by \citet{Terrien} with a precision of $0.14$ dex. \citet{ellie_metallicities} pushed these techniques into the mid-to-late M dwarf regime with spectra of 447 nearby M dwarfs and found that the metallicity of these stars is correlated solely with the equivalent width of the Na line in the near infrared $K$ band to an accuracy of 0.12 dex. Further studies have found that Magnesium and Aluminum features in the near-infrared H-band are good indicators of temperature, radius, and the luminosity of the star \citep{Newton_New}. Attempts to extend this calibration method to the optical regime, where spectral instrumentation is more efficient and more prevalent, have also begun to meet with success. \citet{Mann_early} combined optical and infrared spectra spanning 0.35 - 2.45 microns and found metallicity sensitive features throughout the spectrum. While the infrared features produced a tighter calibration with a precision of $0.1$ dex, an optical-only calibration delivered a precision of $0.16$ dex. \citet{mann} expanded on this optical-only approach (with comparable precision) by bootstrapping the calibration down to later spectral types through early M dwarf - later M dwarf binary pairs. 

Interest in M dwarfs has recently grown due to their attractiveness as targets for planet searches. Their small masses and radii increase the signal amplitudes from both radial velocity (RV) and transit measurements, making detection and detailed characterization of any discovered planetary system easier and accessible from the ground. A number of current and future planet hunting surveys are exclusively focused on M dwarfs or spend a significant amount of observing time on low mass stars. MEarth is an ongoing photometric survey of mid-to-late M dwarfs in the solar neighborhood (Distance, $D \lesssim 33$ pc), looking for low mass rocky planets whose periods may extend into the habitable zone of their star \citep{Nutzman,Berta_2013,jonathan_cool_stars}. The MEarth-North array in Arizona has been in operation since 2008, while a copy of the array in the southern hemisphere, MEarth-South, began operations in early 2014. MEarth is, by necessity, a targeted survey since the nearby M dwarfs are distributed uniformly over the sky. There is typically one target per observing field, in sharp contrast to wide field transit surveys, which observe every star in the field of view. By the nature of this design, properly characterizing targets for prioritization is essential for the survey. 

When preferentially selecting low mass stars as targets for planet surveys, other astrophysical factors may become important in assessing the probability that any individual star hosts a planetary system. The metallicity of the star may be important effects in the planet formation process. 
\citet{Fischer_PM_cor} found that a simple power law in metallicity is sufficient to describe the probability that a star hosts a close and massive planetary companion, and that there is no indication of the migration process of extrasolar planets polluting their host star's atmosphere with metals. It is unclear whether this planet-metallicity correlation extends down to smaller mass planets. \citet{SL_occurence} suggested that a planet-metallicity correlation may be present for small planets around K dwarf stars, but this was later explained by \citet{mann_giants} as contamination in their sample by giant stars. Studies of $Kepler$ stars hosting short period transiting planets have indicated that the existence of large planets in close-in orbits is correlated with stellar metallicity but the existence of small planets is not correlated with metallicity \citep{Buch12,Everett13,Buch14}. However, recent work from \citet{unipm} has again questioned this conclusion, suggesting that there is a universal planet-metallicity correlation across all sizes and all metallicity. \citet{Planet_Metal_New_2015} also investigates this question and finds that stellar metallicity does not affect the radius distribution of planets that form in a system. 

Investigating any putative planet-metallicity correlation down into the M dwarf regime requires a reliable metallicity measurement for a statistically significant number of M dwarfs obtained in a uniform manner, whether that be spectroscopic or photometric. A well-calibrated metallicity measurement of different elemental species assessed directly from spectral measurements themselves is ideal but would require substantive amounts of observing time. Photometric methods lend themselves to larger sample sizes but are hindered by the need for an accurate distance measurement through trigonometric parallax combined with a well determined optical - infrared color. Unfortunately, the nearby M-dwarfs typically lack one or more of these data products. Very few of these stars had accurate distance measurements, as these stars are too dim to have been measured by $Hipparcos$ \citep{Hipparcos_original}. Recently, we have published trigonometric distances to approximately 1500 northern hemisphere M dwarfs from the MEarth survey \citep{2014ApJ...784..156D}. Accurate distance determinations to nearby small stars are still incomplete, particularly in the southern hemisphere, but will soon be much improved by the GAIA mission, now operating. Generally, these studies have broadly eliminated the bottleneck of accurate distance measurements preventing further development of photometric calibration techniques. However, many of the solar neighborhood M dwarfs have only optical magnitudes from photographic plates, with uncertainties of $20\%$ or more \citep{Lepine_33pc_sample}.

In this paper we will describe a photometric calibration of the $MEarth$ bandpass, allowing us to have a large sample of M dwarfs with optical magnitudes measured in a uniform manner. We combine these magnitudes with a compilation of direct distance determinations and spectroscopic measurements to calibrate a color-magnitude-metallicity relation. 
This relation is valid for stars between $M = 0.10M_\sun$ and $M = 0.36 M_\sun$ and has a precision comparable to that of spectroscopic techniques. 
 We release a photometric metallicity estimate for 1507 M dwarf stars in the Northern hemisphere observed with the MEarth survey, and compare the metallicities of the population of nearby M dwarfs with that of nearby G dwarfs. In section 2, we describe the MEarth observatory, target list, and data collection strategy. Section 3 details our data analysis method, photometric model, and our calibration of a photometric metallicity relation.  In section 4, we interpret out results for a volume limited sample of the solar neighborhood. We conclude with brief remarks about future work.

\section{Observations}
While the MEarth-South observatory has been in operation since early 2014, we are considering the data only from the northern array for this work. The MEarth-North Observatory is an array consisting of 8 identical f/9 40 cm Ritchey-Chr\'etien telescopes on German equatorial mounts at the Fred Lawrence Whipple Observatory on Mount Hopkins, Arizona. The telescopes are controlled robotically and collect data every clear night from September through July. The facility is closed every August for the summer monsoons. Each telescope contains a 2048 $\times$ 2048 pixel CCD with a pixel scale of $\approx0.76\arcsec$ / pixel and a Schott RG715 glass filter with anti-reflection coating. The CCD is an e2v CCD42-40 back-illuminated, midband coated AIMO device. The Schott RG715 glass filter consists of two 1.5 mm thick pieces of glass. We provide a quantum efficiency tabulation for our CCD and the transmission for our filter in Table \ref{filter_table}. We note that this tabulation is for a CCD temperature of -20 degrees Celsius, and that there should be a small difference from the operating temperature of $-30^{\circ}$, which we have not measured. The effect of the telescopes' mirror on the overall quantum efficiency is also negligible. While an aluminum coating would vary in reflectivity by approximately 20\% over our bandpass, we are using a customized enhanced aluminum coating from Spectrum Coatings (Deltona, FL), with variations expected to be less than $1\%$.

While MEarth-North has been in operation since 2008, in the 2010-2011 season we replaced our Schott RG715 glass filter with a custom filter similar to a Cousins I filter, in the hope that it would reduce noise from changes in the precipitable water vapor. However, this filter setup did not decrease the noise in our observations, so we switched back to the Schott RG715 filter for the following years. Because of this instrumental change, we will be using data from only the 2011-2014 seasons for the purposes of the absolute photometry presented here.

The MEarth cameras are operated at $-30^{\circ}$ C, and before each exposure we adopt a pre-flash of the detector. While this increases the dark current (which we subsequently subtract off), it suppresses persistence from the previous exposure. We do not detect any discernible effects of residual persistence on the MEarth photometry, and we do not believe it to be significant. Each night we gather flat field frames at dawn and at dusk. Since the MEarth telescopes use German equatorial mounts, we must rotate the telescope by 180 degrees relative to the sky when crossing the meridian. We take advantage of this for flat fielding by obtaining adjacent pairs of flat field images on opposite sides of the meridian to average out large-scale illumination gradients from the Sun and the Moon. Scattered light concentrates in the center of our field of view where our target is typically located. The amplitude of this effect is approximately 5\% of the average value of the sky across the CCD. In order to correct for this, we filter out all large scale structure from our combined twilight flat field and use it only to track small scale features such as inter pixel sensitivity and dust shadows. We derive the large scale flat field response from dithered photometry of dense star fields. Due to the longer wavelength to which our bandpass extends, we must also correct for fringing on the CCD. We find that this effect has an amplitude of $3\%$ across the field of view of the CCD. Fringe frames are gathered by taking long exposures of sparsely populated fields and co-adding them together. 

We measure the non-linearity of our detectors using a dedicated sequence of dome flats. At all count levels, the MEarth CCDs have a slightly nonlinear conversion of photoelectrons to data number, increasing from 1\%-2\% at half the detector full well to 3\%-4\% near saturation. In order to perform accurate photometry, this effect needs to be minimized and corrected. The MEarth target stars are generally a different magnitude than the comparison field stars. The MEarth pipeline has been developed to minimize non-linear CCD effects as part of our planet search program. Our exposure times are set to avoid surpassing 50\% of our detector's full well in order to minimize the nonlinearity effect, and our nonlinearity measurements are used to correct for the remainder of this effect. After these corrections we see no evidence that nonlinearity plays a significant role in our photometric performance. We also correct for varying exposure time across our field of view due to shutter travel time. Due to the low exposure times needed to measure this effect, we measure the exposure time correction at low exposure times with a series of short dome flats and then measure the shutter pattern with a series of twilight flats. Using twilight flats reduces additional error from linearity correction when deriving the shutter pattern. We find the linearity correction and fringing pattern to be stable. During the summer monsoon, which occurs each year in August, data acquisition is halted and the telescopes are shut down. This time is used to perform maintenance activities on the telescopes. 

We measure stellar positions using a method similar to that of \citet{I85}. We estimate local sky background by binning each image into 64$\times$64 pixel blocks, and measure the peak of the histogram of the intensity of these pixels. We then interpolate this lower-resolution map to measure the background level anywhere in the image. Stellar locations are measured using intensity weighted first moments (the centroid of the star), and the total flux is measured using an 8 pixel ($\approx 6.2$ arcsecond) aperture radius in an effort to closely match the measurements done by the ongoing Landolt standard stars project (see: \citealt{Landolt13b} and references therein). Pixels that lie partially within this circular aperture are weighted according to the fraction of that pixel that lies within the ideal circular aperture. Sky background is estimated with a sky annulus between 24 and 32 pixels (18.6 - 24.8 arcseconds) away from the stellar photo center. We also adopt an aperture correction \citep{aperture_correction} in order to correct for the flux that falls outside of our aperture. The typical size of this correction is 0.04 magnitudes.

The MEarth-North target list consists of 1844 nearby M dwarfs selected from \citet{Lepine_33pc_sample}, a subset of the LSPM-North catalog \citep{Lepine_catalog} believed to be within 33 pc of the Sun and with stellar radius $R < 0.33 R_\sun$ \citep{Nutzman}. These targets are uniformly distributed across the Northern sky ($\delta > 0^{\circ}$) such that typically only one target exists in the MEarth 26\arcmin x 26\arcmin$ $  field of view, with the exception of multiple systems and occasional unrelated asterisms. Each field is targeted individually, and is selected to contain sufficient comparison stars to obtain high precision relative photometry and astrometry.

For each observing season (September through July), MEarth gathers data at two different cadences. All targets are observed at approximately $10$ day intervals (when they are visible) as part of the MEarth astrometric program, for which we recently published trigonometric parallaxes to a majority of our targets \citep{2014ApJ...784..156D}. These observations take place at a variety of air masses, depending on when the scheduler can fit an observation between planet search target observations. For nights where the object is visible on both sides of the meridian, we sometimes schedule two astrometric data points in one night in order to resolve degeneracies in our parallax model (see \citealt{2014ApJ...784..156D} for details regarding our astrometric reduction pipeline). 

A subset of these targets is selected for our planet search campaign, and is observed each night. Within that night, our planet search targets are observed at a 20-30 minute cadence, ensuring that we obtain at least two in-transit data points for a typical one hour duration transit. Since we are monitoring these stars for planets, observations are taken at many different air masses throughout the night. We discuss how we take advantage of this for our photometric analysis in section 3. For the results presented here, we have utilized all of the MEarth data taken at both astrometric and planet search cadence.

\begin{figure}
\centering
\includegraphics[scale=0.75]{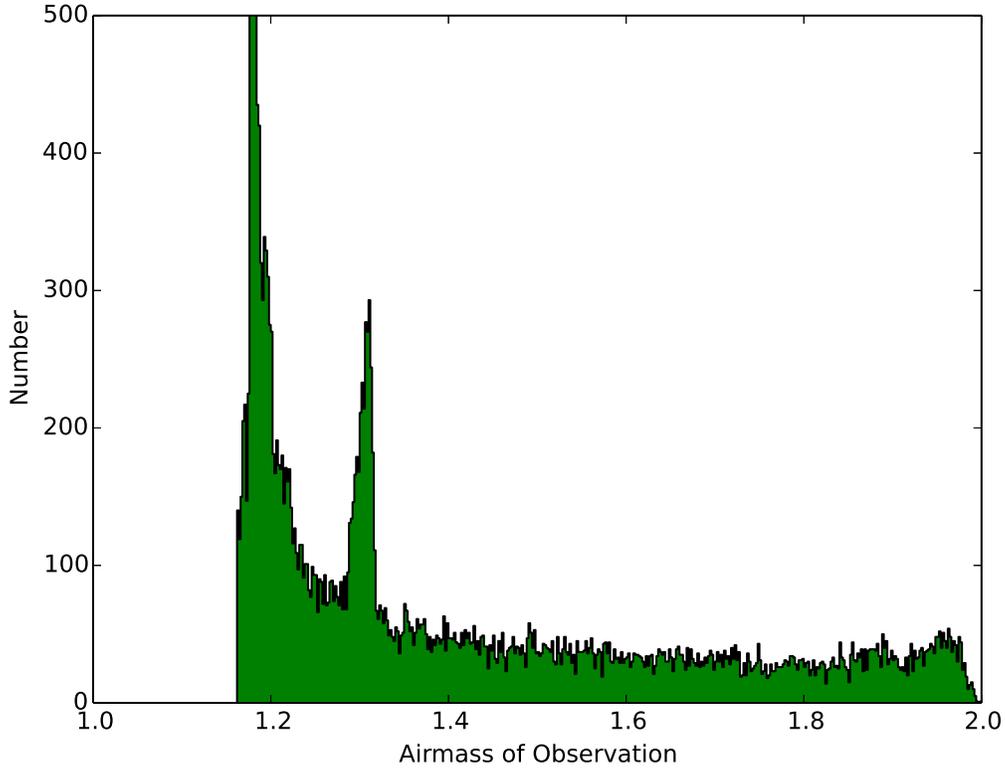} \\
\caption{Histogram of the observed air masses for all of the Landolt standard fields observed across all telescopes during the course of the survey. The distribution is very non-uniform and does not reflect the time each standard field spends at any given airmass. The dearth of observations at low airmass is due to the latitude difference between Mt. Hopkins, Arizona and the Landolt equatorial standard fields observed in our program.  Our scheduling program utilizes a zenith distance threshold to avoid exclusively observing Landolt fields at high airmass as soon as they rise, which creates the secondary peak in the histogram at 1.3 air masses.}
\label{standard_star_airmasses}
\end{figure}

In addition to our target observations, each night we observe Landolt standard star fields \citep{Landolt92,Landolt09,Landolt13b,Landolt13a}. We typically observe 3-4 standard star fields per telescope in a given night and each field is observed only once in any given night in order to minimize its impact on the main survey. Standard fields tend to be taken at the beginning of the night, and at higher airmass than target observations. We show a histogram of the airmass for all standard field observations across all telescopes in Figure \ref{standard_star_airmasses}. The absence of observations at low airmass is due to the difference in latitude and declination between our observatory and the equatorial standard star fields. The secondary peak at an airmass of 1.3 is a feature of our scheduling system. To avoid exclusively taking Landolt standard observations at high airmass, we use a zenith distance threshold for fields that are up for significant portions of the night to prefer lower air masses. Aside from this, it is readily apparent that while we obtain observations of standard star fields at all air masses, the distribution of observed air masses does not reflect the relative amount of time that any given standard field is visible at that airmass. However, we do not believe that this has any significant effects on our derived photometry.

\section{Analysis}
All data (long cadence, calibration fields, and short cadence) were reduced using the MEarth data reduction pipeline (see \citealt{berta2012} for an in-depth description of the MEarth data reduction procedure). In order to calibrate MEarth's measured photometry onto an absolute scale, we must make corrections for the effects of the atmosphere and light loss in the telescope. These effects are both color dependent and airmass dependent. We adopt the following model:

\begin{equation}
m_{*,MEarth} = z_i - 2.5 \log_{10}(f_{obs}/t_{exp}) - \mbox{AC} - \biggl( k_{1,i} + k_{2,i} (m_{*,MEarth} - m_{*,K}) \biggl) \times (X-1)
\label{equation1}
\end{equation}

where $m_{*,MEarth}$ is the magnitude of the star in the $MEarth$ band, $m_{*,K}$ is the magnitude of the star in the 2MASS $K_s$ band, $z_i$ is the zero point of the telescope, $f_{obs}$ is the observed flux (in counts) for the star, $t_{exp}$ is the exposure time of the image, AC is the aperture correction for the image, $k_{1}$ and $k_{2}$ are fitted coefficients to correct for airmass and color effects, respectively, and $X$ is the airmass of the object. $i$ denotes the night that the observation was taken. 

While the standard field observations alone are too few in number on any given night to provide simultaneous constraints in $k_1$ and $k_2$, we utilize all the stars of sufficient brightness in every target field to also inform our fits. Traditionally, standard star fields are tracked across the sky and periodically revisited during the night, allowing multiple measurements of the same standard stars across a range of airmasses. However, since we are calibrating the $MEarth$ photometric system itself, and utilizing $MEarth$-$K_s$ color in our nightly fits, we can utilize {\em every} observed star with a unique 2MASS identifier to constrain the airmass and color coefficients of our fits. Due to the sheer number of these stars compared to the number of Landolt standards, these stars dominate the fit for these parameters in any given night. This method allows us to benefit from sampling a wide range of airmasses continually throughout the night rather than a handful of discrete pointings from the Landolt standard field observations. The zero point of each telescope, however, is uniquely constrained by the standard field observations. We use reference stars that are between $MEarth$ magnitudes 9.0 and 13.0, which have sufficient signal to noise (S/N) ratios without being saturated in the image. 

We fit our observations iteratively, first determining $k_1$ and $k_2$ for each individual night using all the data that were taken on that night, and then fitting for the zero point ($z_i$) of the telescope using solely the standard field observations. While our filter is different from the filters used by \citet{Landolt92,Landolt09,Landolt13b} and \citet{Landolt13a}, we base our zero point correction to most closely match the Landolt {\em I} measurements for these stars at $MEarth$-$K_s$ color = 0. We reduce the data from each telescope independently from the data from other telescopes, in order to avoid systematic color effects that may be present between the different telescopes. However, we have recently changed MEarth's observing strategy such that observe target stars on two telescopes to increase our signal to noise per star and mitigate systematic effects in our search for small transiting extra-solar planets \citep{Berta_2013}. For stars that have been observed in this mode, we are able to compare independent measurements of the broadband $MEarth$ magnitude and assess possible systematic effects with our data (see section \ref{Error_section}). 

We remove non-photometric data by discarding a night if the standard deviation of the residuals across all images taken that night is greater than 0.02 magnitudes. After this cut, each telescope has between 100 and 200 photometric nights of data, depending on whether each telescope was operating or being repaired on any given night deemed to be photometric. On any photometric night, each telescope obtains approximately 10,000 individual measurements of stellar fluxes, across the M-dwarf target stars, comparison stars in the field, and the Landolt fields. In Table \ref{coefficient_table}, we list the color and airmass coefficients as well as the zero points for each telescope on each night that was deemed photometric, while in Table \ref{standard_table} we list the $MEarth$ magnitude determined for each Landolt standard star observed.

\subsection{The $MEarth$ optical magnitude for 1802 M dwarfs}
\label{Error_section}
We apply our derived coefficients to all of our available data points for each MEarth target, estimating a $MEarth$ band magnitude using equation \ref{equation1}. We report the $MEarth$ magnitude as the median of all of our measurements across all photometric nights and all observing seasons. We estimate our error as the median absolute deviation (MAD) from this median value, and adjust it to $1$ Gaussian $\sigma$ by multiplying by $1.48$ \citep{Stat_book}. This method ensures that single erroneous data points, which could be contaminated from cosmic rays in the image, flares from the M-dwarf itself, or other reasons from significant skewing our measurement. We note that for some of our stars the reported uncertainty in the $MEarth$ magnitude is greater than the standard deviation required for any individual night to be considered photometric in our analysis. This is likely an astrophysical effect, as many M-dwarfs stars are variable in time due to stellar activity and rotation (starspots rotating in and out of view \citealt{Irwin_rotation}), and we do not consider this to be an indication of significant systematic errors associated with our measurements. 

\begin{figure}
\centering
\includegraphics[scale=0.5]{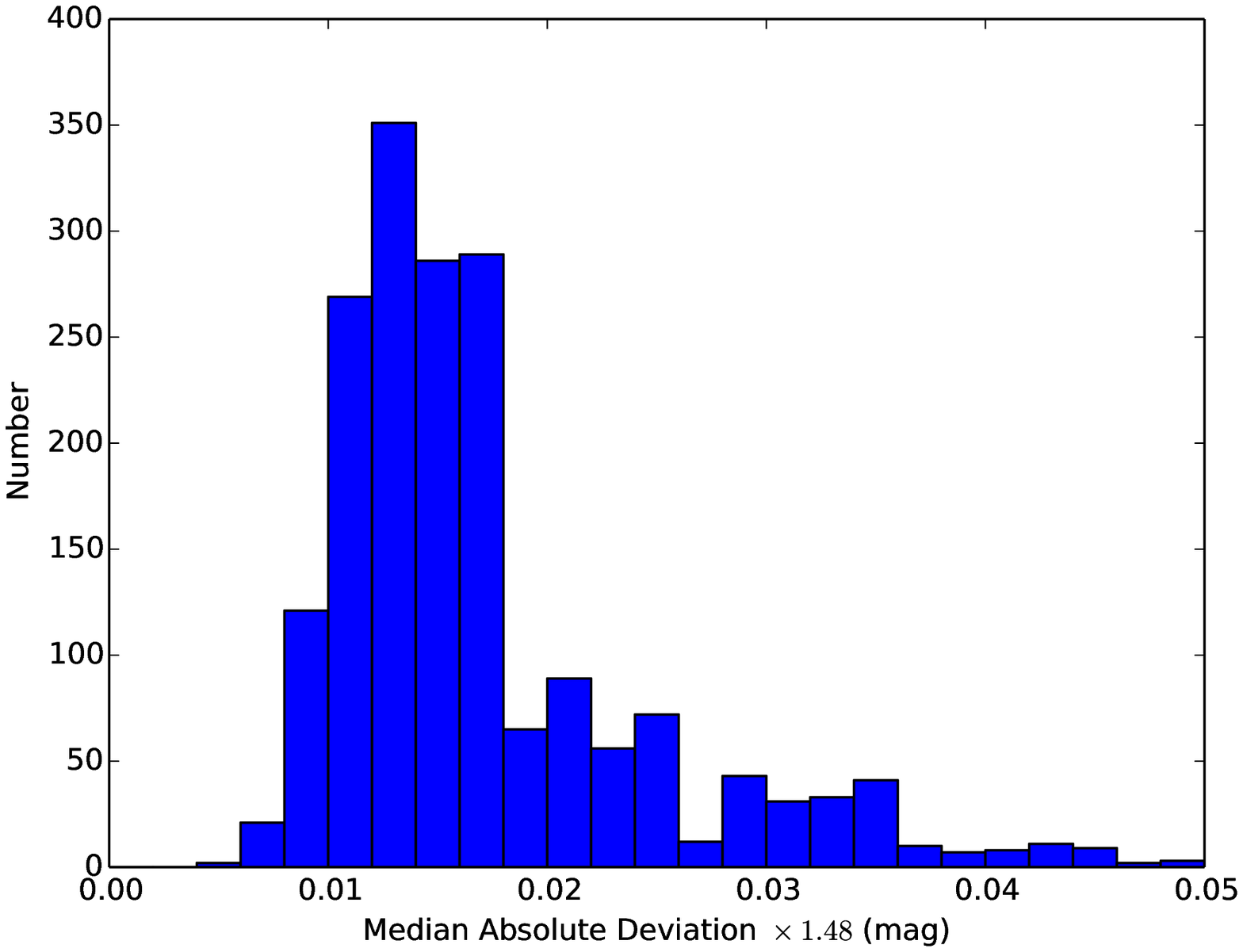} \\
\includegraphics[scale=0.5]{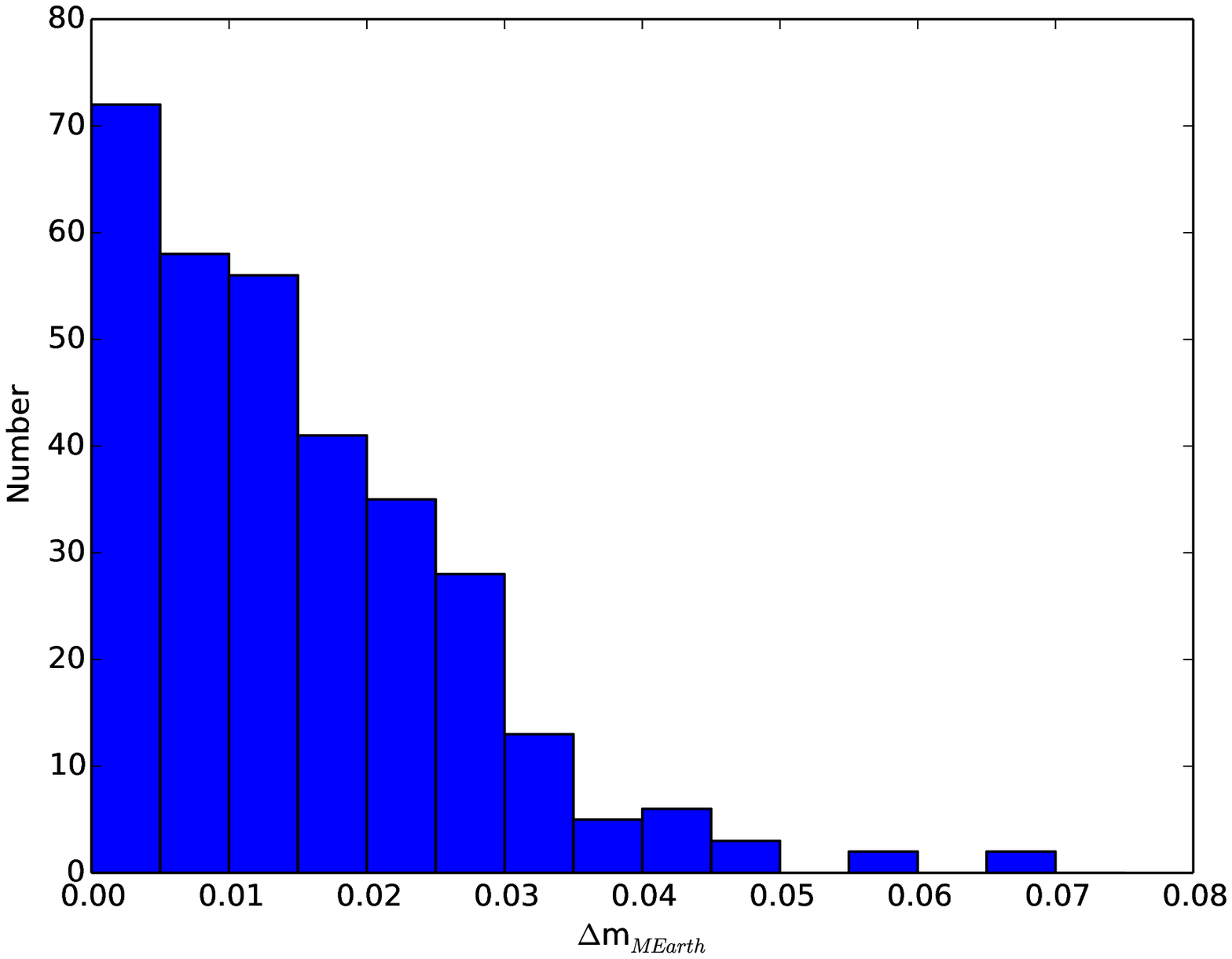} \\
\caption{Top: Histogram of the internal errors of the $MEarth$ magnitude for each target star. Typical errors are 0.015 magnitudes, and are derived from the absolute deviation from the median of all data points for the target star, adjusted to $1\sigma$. Bottom: Histogram of the difference between the $MEarth$ magnitude measured for 322 of our target stars that have measurements taken with multiple MEarth telescopes. The 68th percentile of the absolute difference between the magnitude measurements between telescopes for the same star is 0.018 magnitudes, approximately the typical error for stars with only single telescope measurements. Therefore, we do not believe that there are any significant systematic effects between telescopes and that they all are calibrated to the same magnitude system. The median number of data points a star has is 123, and less than $5\%$ of our sample has less than 30 photometric observations.}
\label{telescope_comparison}
\end{figure}

The median uncertainty in our $MEarth$ magnitudes is 0.015 magnitudes, or $1.5\%$. We show a histogram of these errors in Figure \ref{telescope_comparison}. A small number of our targets have errors bars significantly larger than this typical value. $143$ of $1802$ stars have a error bar greater than 0.03 magnitudes, and 9 stars have an error greater than 0.05 magnitudes. Extremely large uncertainties are typically due to contamination of our target flux with that of a brighter companion. We indicate whether a star may be potentially contaminated in Table \ref{big_table} (see below). We remove blended stars from our table which we believe to be so significantly blended from a particularly close and bright companion that we do not believe our measurements accurately reflect either the combined magnitude or the magnitude of one of the components. 

MEarth has begun observing our target stars with two telescopes in order to increase our signal to noise and sensitivity to small planets \citep{Berta_2013}. This change in survey strategy allows us to independently measure the $MEarth$ magnitude of these stars and assess possible sources of systematic error in our absolute photometric calibration. 322 of our stars have measurements with multiple MEarth telescopes, and we show a histogram of the difference between these measurements in Figure \ref{telescope_comparison}. The 68th percentile of the absolute difference between the magnitude measurement of the same star measured from different telescopes is 0.018 magnitudes, approximately the same as the internal error determined from individual telescopes. Therefore, we conclude that each MEarth telescope is calibrated to the same photometric system and that there are no additional uncertainties accrued by combining the magnitude measurements from multiple telescopes. We proceed with this assumption for the rest of this paper. 

We list the optical magnitude of each star in Table \ref{big_table}. We also provide three blend flags to indicate whether the magnitude measured for this star may have significant contribution from either a binary companion, nearby bright star, or a star lying underneath the aperture for the target star. Our first blend flag indicates whether the aperture contains any additional stars other than the target star. This includes background objects as well as close common proper motion pairs. In many cases, this contamination is negligible, but we recommend assessing the importance of the contaminating flux using the MEarth finder charts available on the MEarth website (see below). Our second blend flag is for objects that have two individual entries in the LSPM-North catalog but are unresolved in the MEarth telescopes. Additionally, some of these objects are also unresolved in 2MASS. Our third blend flag indicates a known blended object that also has only a single entry in the LSPM-North catalog. For these sources, we also provide a reference. These sources include visual binaries split in guider-camera observations at IRTF as well as spectroscopic binaries and those identified via imaging. Finder charts for these systems are available on the MEarth website\footnote{https://www.cfa.harvard.edu/MEarth/Welcome.html}. These finder charts include plate scans from the Digitized Sky Survey, which when combined with the high proper motion nature of our stars, allow us to see whether there is significant contaminating stellar flux at the current position of the M dwarf. 

\subsection{A color-magnitude-metallicity relation anchored by spectroscopic measurements}
Past studies have indicated that the location of M-dwarfs on the ($V$-$K_s$) \textemdash $M_K$ color magnitude diagram is significantly affected by the star's metallicity \citep{B05,johnjohn_metallicity,Neves12,Schlaufman_Laughlin_12}. Since an M-dwarf's absolute K$_s$ band magnitude seems to be almost entirely determined by the star's mass \citep{Delfosse}, the metallicity dependence originates primarily in the optical magnitude of the star. The cooler temperature in these stars' atmospheres allow the formation of many molecules with broad (and overlapping) absorption bands that act to suppress the optical flux, creating this effect. By combining existing distance measurements and spectroscopic metallicity measurements with our $MEarth$ magnitude measurements, we can create a classical color-magnitude diagram using $MEarth$ as the optical component of the color.

\begin{figure}
\centering
\includegraphics[scale=0.5]{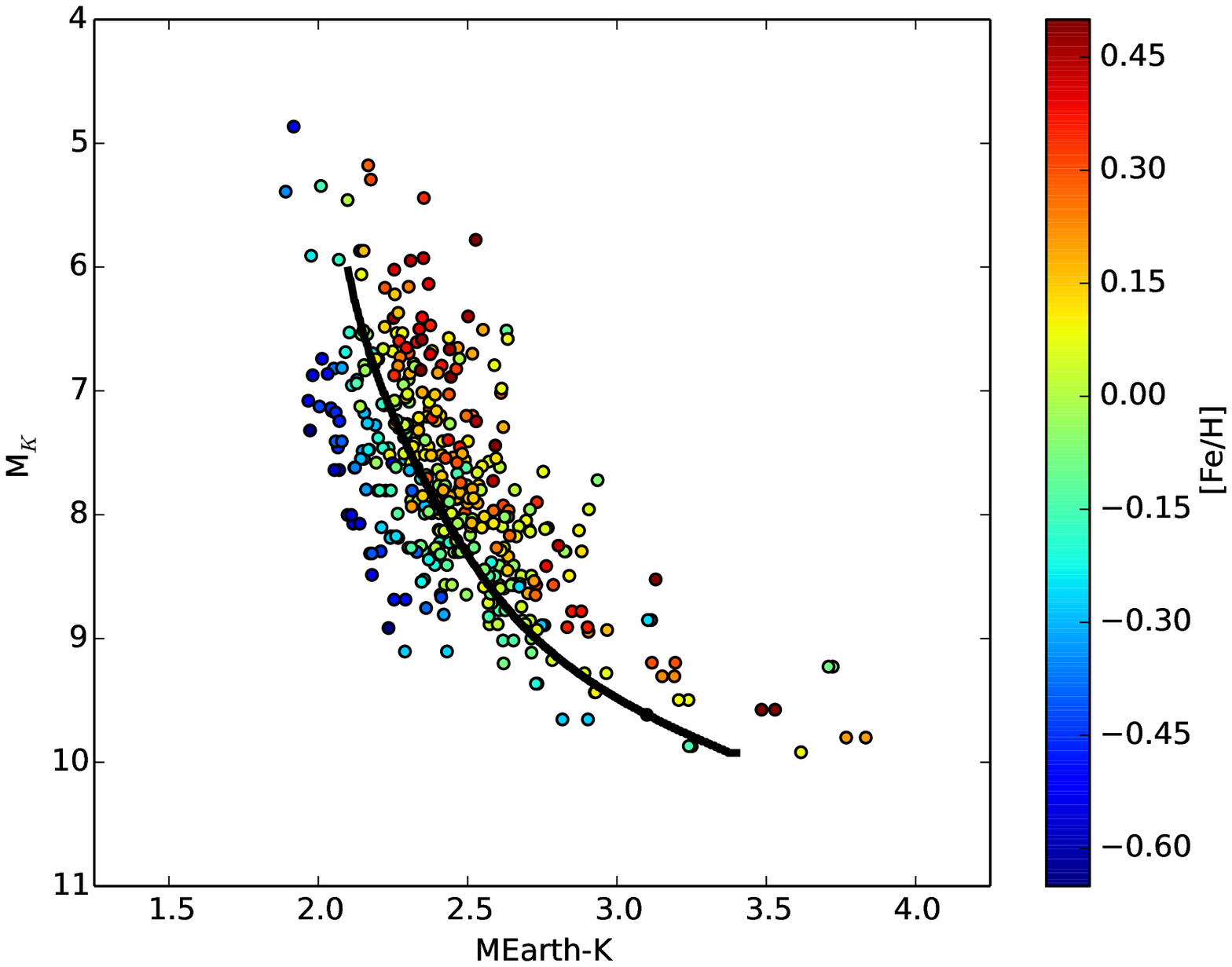} \\
\includegraphics[scale=0.5]{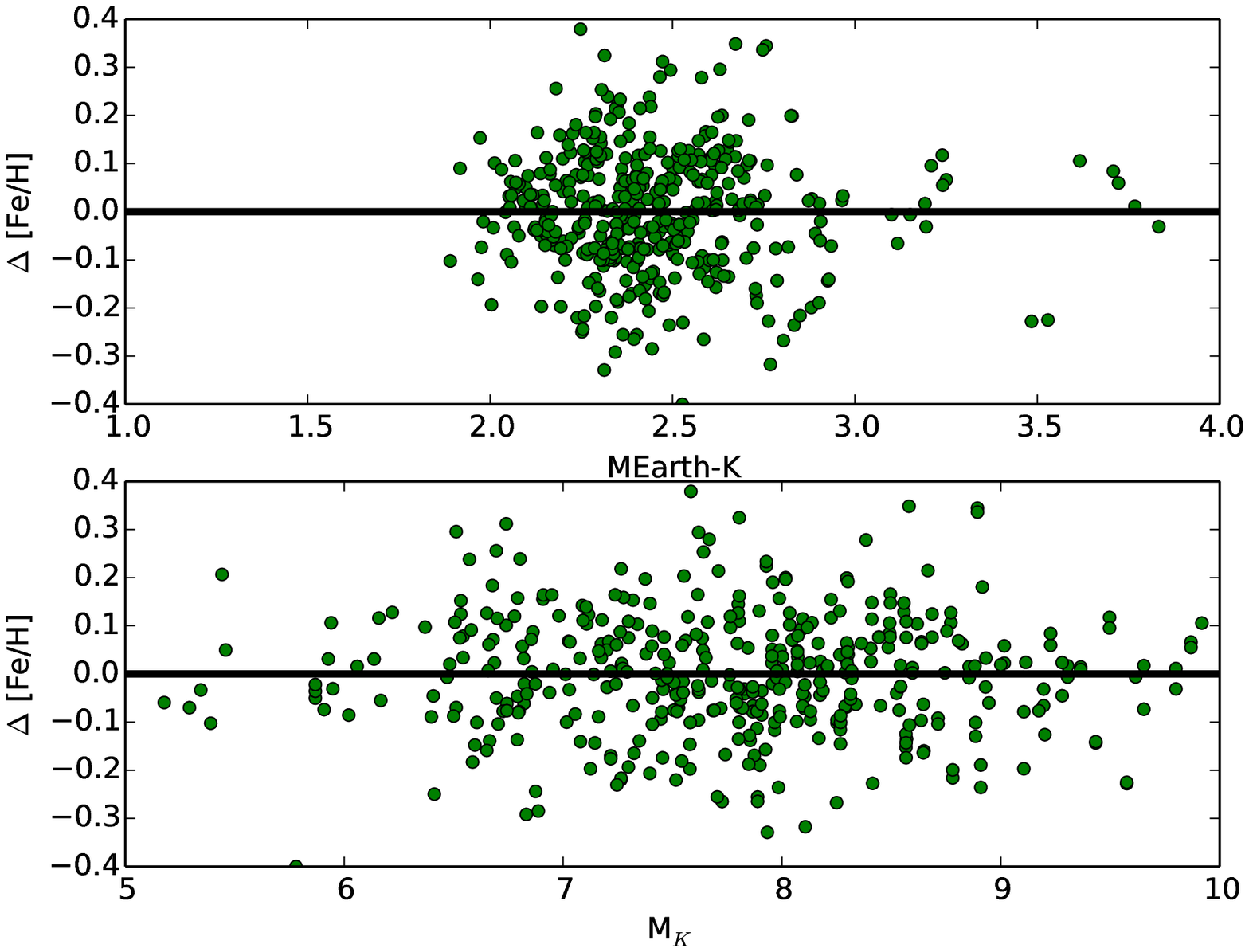} \\
\caption{Top: Absolute K band magnitude vs $MEarth$-$K_s$ color for all stars that have a trigonometric parallax and an infrared spectroscopic measurement. The color scale is the infrared spectroscopic measurement. While $M_K$ is an excellent mass indicator \citep{Delfosse}, we find that the $MEarth$-$K_s$ color of an M dwarf is very highly correlated with the metallicity of the object. Increased metallicity results in a redder color due to line blanketing at optical wavelengths. The black line is our interpolated main sequence at [Fe/H] = 0.0. \newline
Bottom: Residuals in our 2 dimensional metallicity interpolation as a function of $MEarth$-$K_s$ color and $M_K$. We find no significant systematic trends. The standard deviation of our 2D interpolation is $0.11$ dex, comparable in precision to the infrared spectroscopic metallicity indicators. }
\label{color_magintude}
\end{figure}

\begin{figure}
\centering
\includegraphics[scale=0.75]{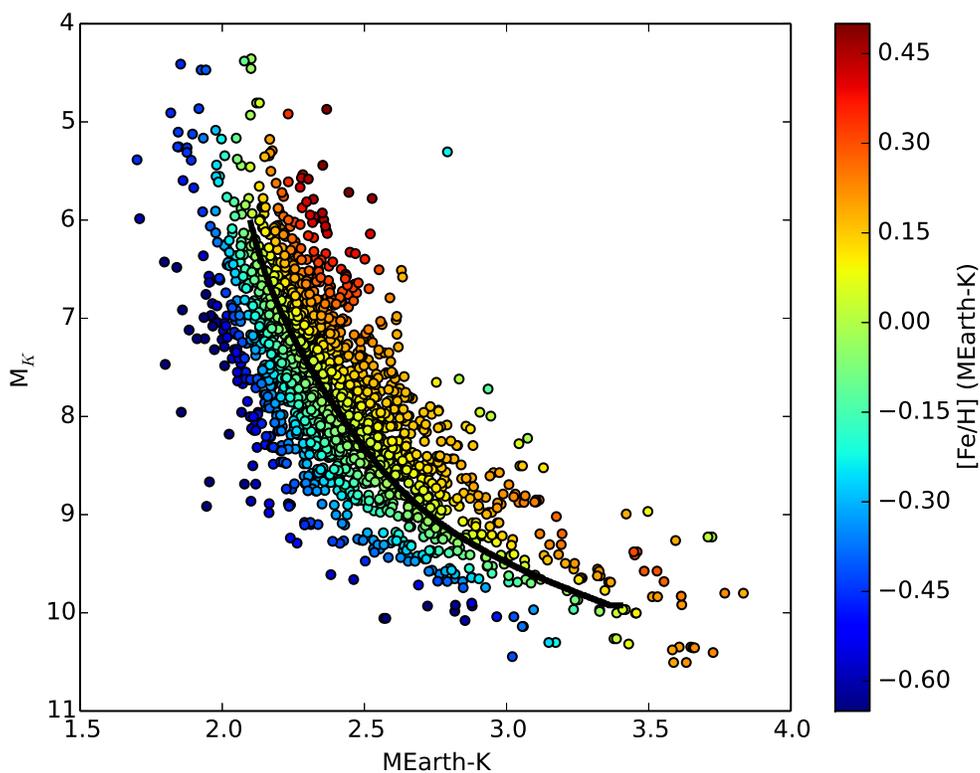} \\
\caption{Same as for the top panel Figure \ref{color_magintude} but for the entire MEarth sample and utilizing our 2-dimensional photometric metallicity interpolation instead of independent spectroscopic metallicity results. The black line is a solar metallicity main sequence. Our work increases the number of nearby M dwarfs with metallicity measurements, and provides a statistically significant sample measured in a uniform manner.}
\label{full_sample_diagram}
\end{figure}

In Figure \ref{color_magintude}, we plot the color-magnitude diagram for the 307 stars for which we have a trigonometric distance measurement and a near-infrared spectrum from \citet{ellie_metallicities}. For stars with multiple trigonometric distance determinations we utilize the observation with the lowest formal reported error bar instead of averaging all observations. Metallicity measurements based on the NIR spectra are taken from \citet{Newton_New}, who combined the metallicity calibrations of \citet{Mann_early} for early type M-dwarfs and \citet{mann} for later type M-dwarfs. The more metal rich stars at a given M$_K$ are redder than those that are more metal poor, as has been noted in previous studies \citep{B05,Neves12,Schlaufman_Laughlin_12}.  We have fit a smoothed 2d spline to this data using the method described by \citet{spline_fitting} and show the [Fe/H] = 0 main sequence in Figures \ref{color_magintude} and \ref{full_sample_diagram}. Color-magnitude main sequences at the mean solar neighborhood metallicity of [Fe/H] = -0.1 and at solar metallicity ([Fe/H] = 0.0) are available in Table \ref{main_sequence_track}.

The dispersion in metallicity from this interpolation is $0.11$ dex, equal to the precision derived from the spectroscopic measures themselves. We note that as this method is calibrated to existing infrared spectroscopic metallicity determinations, it inherits any systematic errors associated with these measurements. We find no significant systematic trend with the absolute K-band magnitude of the star or with the $MEarth$-$K_s$ color of the star. We note that the majority of our sample lies between $M_K = 6.5$ and $M_K = 9.0$ (between 0.11$M_\sun$ and 0.36$M_\sun$ using \citealt{Delfosse}). Therefore, this is the domain over which our results are the most robust and well-constrained. The higher mass M dwarfs included in the calibration are mostly stars who had inaccurate optical colors and whose parameters were inaccurately estimated, before we obtained trigonometric parallaxes for most of these targets. Our attempt to exclude these higher mass stars using the color-based estimates is likely to introduce a bias that systematically excludes bluer (more metal poor) stars for masses (as estimated using the more reliable $M_K$-based calibration) above $0.31 M_\sun$. The MEarth sample also has few targets with $M_K > 9.0$. Therefore the interpolation is not as strongly constrained in this regime. While we report metallicity estimates for stars in these magnitude ranges, they should be treated with more caution than the results reported for stars with $6.5 \leq M_K \leq 9.0$. We present our metallicity estimates, along with our optical magnitude measurements, for our full sample in Table \ref{big_table}. There are 277 stars in our sample that do not have a trigonometric distance estimate. For these stars, we fix the star's metallicity at [Fe/H] = -0.1 and provide an estimate of the distance modulus ($m-M$) of the star. 

We show the color-magnitude diagram for all MEarth targets with a determined magnitude and trigonometric parallax, using the metallicities inferred from our relation, in Figure \ref{full_sample_diagram}. We note that 29 stars are much redder than what is expected from their absolute K-band magnitude and lie significantly outside the calibration range for our spline. This far from our calibrated sequence our calibration turns-over and our method assigns them unrealistically low metallicities. An unresolved equal-mass binary would be shifted upwards by 0.75 magnitudes in Figure \ref{full_sample_diagram}. An unresolved binary of unequal mass will be shifted upwards (due to the extra luminosity) but also to redder colors (as the the average color of pair is shifted). Undoubtedly our sample must contain a significant number of unknown, unresolved binaries. We suggest that these redder or over luminous stars that lie outside of our interpolation regime may be unresolved multiples. For each star, if our metallicity interpolation predicts a higher metallicity for a hypothetical star at the same absolute magnitude but 0.25 magnitudes bluer, then we flag this as an over luminous source in Table \ref{big_table}.

There are 894 stars in our sample that have spectral types determined from optical spectra by various groups (see Table \ref{big_table} for spectral types and sources). These spectral types are compiled from the literature and, in cases where there are more than one spectral type measurement, we prioritize spectral typing methods involving direct comparison to spectral standards \citep{Lepine_Spec}. In some cases, the spectral type is derived from measurements of molecular band indices. We do not average spectral types from multiple sources, but use the spectral type from the highest priority source. 

\begin{figure}
\centering
\includegraphics[scale=0.75]{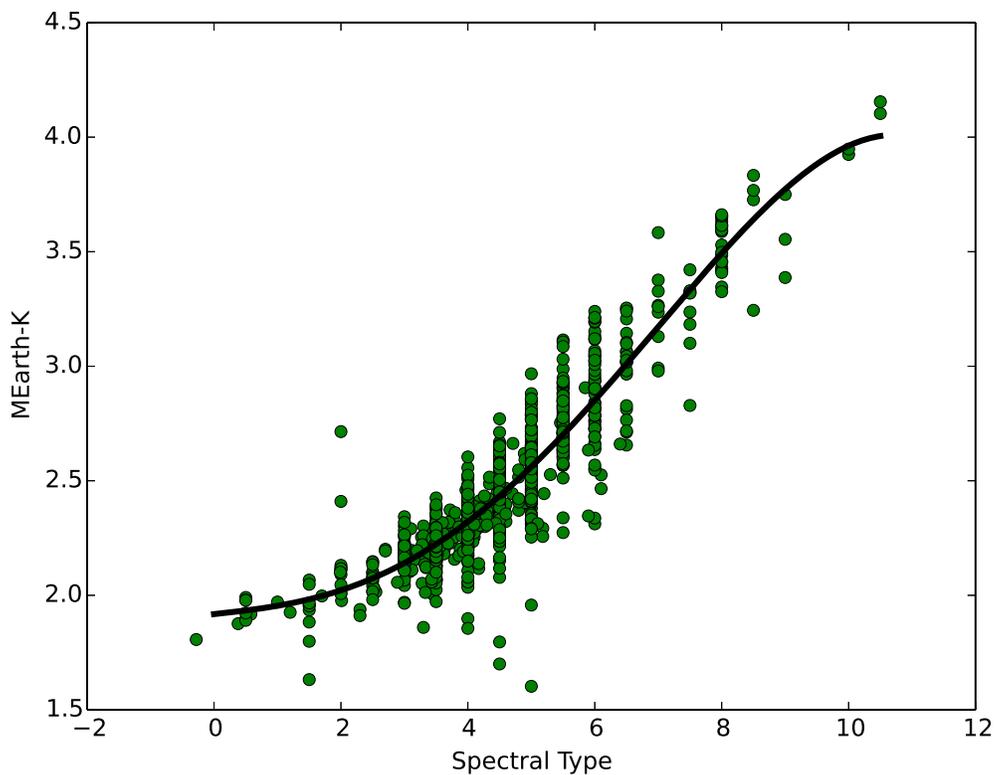} \\
\caption{Evolution of the $MEarth$-$K_s$ color as a function of M dwarf spectral type with a third order polynomial fit through the sequence (shown as a black line). The star earlier than M0 is a late K dwarfs while M10 and later begins the L dwarf sequence. All spectral types are taken from the literature and may not be determined in a strictly uniform manner. We find that the average $MEarth$-$K_s$ color sequence continuously becomes redder for later spectral types although there is significant amount of variation due to metallicity, which this fit does not take into account.}
\label{spectral_type_figure}
\end{figure}

We have investigated the evolution of the $MEarth$-$K_s$ color of a star as a function of spectral types (see Figure \ref{spectral_type_figure}). We find that a 4th order polynomial is sufficient to capture the evolution of the $MEarth$-$K_s$ color through this spectral range. This polynomial is:

\begin{equation}
\textrm{$MEarth$-K$_s$} = (-4.5\times10^{-4}) \times N^4 + (6.5\times10^{-3}) \times N^3 - (3.2\times10^{-3})\times N^2 + (3.6\times10^{-2}) \times N + 1.9
\end{equation}

where N is the numerical spectral type of the star, $0$ being an M0 type star, $5$ being equivalent to an M5.0,  and $10$ being an L0 dwarf.

 While the $MEarth$-$K_s$ color of a star is a good indicator of spectral type through mid-M spectral sequence, we find that for the earliest M dwarfs the $MEarth$-$K_s$ color does not change substantially. Furthermore, the MEarth sample selection criteria is designed to eliminate early M stars from our sample. Early type M dwarfs that make it into our sample are likely to be systematically redder than typical stars of that spectral type, so that they may masquerade as later type stars when utilizing simple color cuts. Therefore, these stars may not be representative of early M stars in general, and we suggest caution when using this equation for spectral types earlier than M2. At the earliest spectral bins we also have a higher proportion of outliers. These may be systematically affected by contamination if they are close to another star or may be subject to higher systematics than the later spectral types that we are targeting in our observations (greater nonlinearity or saturation). These red outliers are LSPM J1204+7603 and LSPM J2307+1501. LSPM 1204+7603  is flagged as a potential blend with other background stars. We believe that this is the cause for this star being an outlier. The magnitude of LSPM 2307+1501 does not appear to be an outlier in our analysis. We suggest the reported spectral type may be incorrect. 

\section{Discussion}
\subsection{The metallicity of the solar neighborhood M dwarfs}
We present a large sample of uniformly determined metallicity estimates for northern M dwarfs in the Solar Neighborhood. While the utility of this technique is somewhat limited due to the non-standard nature of our bandpass and the requirement of an accurate distance measurement to the star, we can use the measurements for 1567 stars with trigonometric parallax measurements in order to investigate the nature of the nearby M dwarfs.

We restrict ourselves to a volumetrically complete sample to avoid potential biases. \citet{2014ApJ...784..156D} showed that the MEarth sample is complete out to a distance of 20 parsecs within the color, estimated distance, and $R_{est} < 0.33 R_\sun$ sample selection criteria of \citet{Nutzman}. We will restrict ourselves to this volume for this analysis. While \citet{2014ApJ...784..156D} corrected for the subsample of stars culled from the MEarth sample for being too bright for our survey or too close to a bright star, we do not make such a correction here as these missing stars do not have a $MEarth$ magnitude. We do not believe that this will significantly affect our results because there are not many of these stars and they are likely to not be significantly biased in metallicity. Furthermore, the study in \citet{2014ApJ...784..156D} divided the MEarth sample by spectral type, for which completeness can vary as a function of distance. Here we are studying the entire MEarth sample and therefore dilute the effect of a handful of known nearby stars being removed from the analysis.  

\begin{figure}
\centering
\includegraphics[scale=0.75]{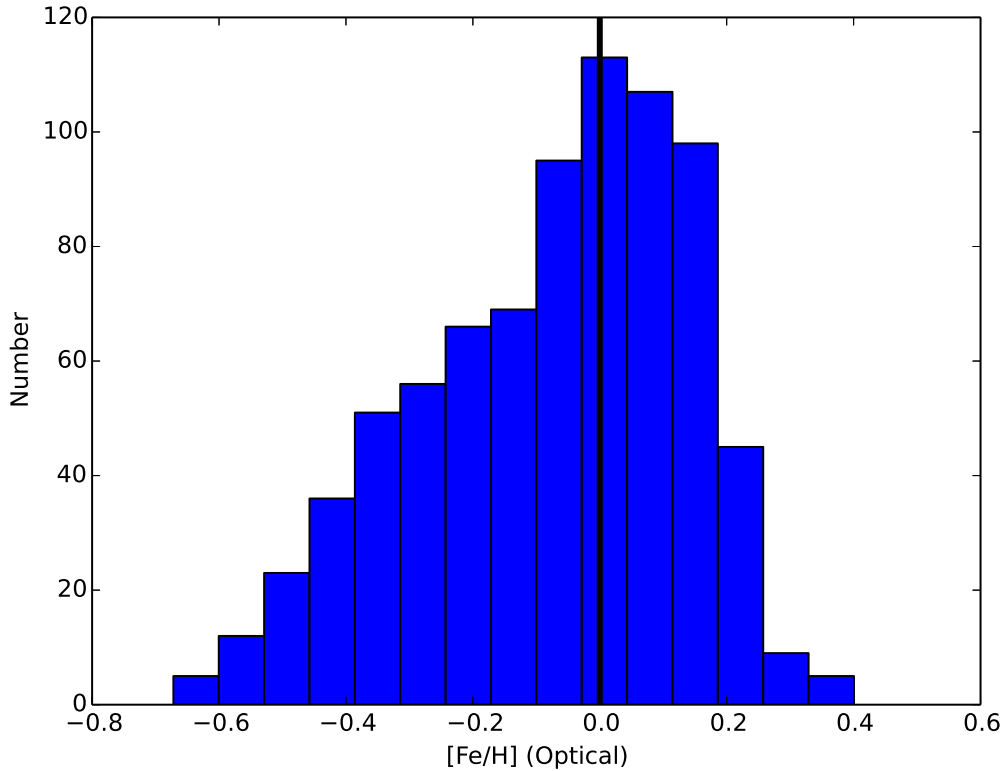} \\
\caption{Histogram of the metallicities of the M dwarfs within 20 parsecs of the Sun in the MEarth sample. The black line at [Fe/H] = 0.0 is the value for the Sun. The median metallicity for these solar neighborhood M dwarfs is [Fe/H] = $-0.03 \pm 0.008$, coincident with the solar metallicity. The asymmetric distribution of metallicities may be partially due to the saturation of lines at high metallicity in the spectroscopic sample used to calibrate our photometric relation. This saturation occurs around [Fe/H] $\approx$ 0.2. If so, then the true metallicity of stars shown here with [Fe/H] $>$ 0.1 may be distributed out to higher metallicity. We find that the metallicity distribution of the solar neighborhood M dwarfs is similar to that of the solar neighborhood G dwarfs.}
\label{histogram}
\end{figure}

In Figure \ref{histogram} we show the metallicity distribution of M-dwarfs in the solar neighborhood within 20 parsecs of the Sun. The median value of these metallicities is [Fe/H] = $-0.03 \pm 0.008$, approximately solar metallicity. The sharp drop beginning at [Fe/H] = 0.15 is due to the saturation of the sodium line used in these metallicity measures \citep{Newton_New}. Since our photometric metallicity calibration is based upon spectroscopic measurements derived from these lines, we may be subject to systematic errors at high metallicity. Therefore, this sharp drop off and lower proportion of stars at high metallicities may not be real, and the excess of stars between [Fe/H] = 0.10 and [Fe/H] = 0.15 is likely to be distributed across larger metallicities.

We avoid this potential source of systematic error by looking  strictly at the relative proportion of stars in our sample that are metal poor. For this sample, 29 of the 565 of our stars within 20 parsecs have a metallicity of [Fe/H] = -0.5 or less, approximately $5\%$. This is comparable to the proportion measured by \citet{Haywood01}, who measured that $4\%$ of their 383 solar neighborhood G dwarfs have a metallicity less than [Fe/H] = $\mathrm{-}0.5$. We cannot reject the null hypothesis that the local M-dwarfs are drawn from the same metallicity distribution of the local G-dwarfs based upon the fraction of metal poor stars. Therefore, they are likely derived from the same star formation history.

\section{Conclusions}
We have calibrated the $MEarth$ photometric system using standard star observations taken routinely as part of general MEarth operations. By combing these results with trigonometric distance measurements and spectroscopically determined metallicity estimates from \citet{ellie_metallicities,Newton_New}, we have calibrated a new color-magnitude-metallicity relation. The location of an M dwarf in $M_K$ vs $MEarth$-$K_s$ space is a metallicity indicator with comparable precision to infrared spectroscopic measurements, and we provide a photometric metallicity estimate for 1567 northern M dwarfs in the solar neighborhood with precisions of $0.1$ dex. We find that the metallicity distribution and its median value are consistent with that of the nearby G dwarfs.

The nearby northern M dwarfs tabulated here are a promising hunting ground for ongoing planet searches like MEarth \citep{Nutzman} and APACHE \citep{APACHE}, and also future surveys like CARMENES  \citep{CARMENES}, the Habitable Planet Finder Spectrograph \citep{HabPlanFin}, SPIROU \citep{SPIROU}, and EXTRA (X. Bonfils, personal communication). Planets found by these endeavors will allow us to investigate correlations between planet occurrence rate and stellar metal abundances. However, in order to do so, we need not only an accurate measure of M dwarf metal content, but also the capability to apply this uniformly on a statistically significant number of stars. The metallicities presented here offer a precision comparable to spectroscopic techniques, but the advantage of being applied over a much larger sample without the need of a significant investment in additional telescope time. Future improvements in this method will be made possible by GAIA\footnote{http://sci.esa.int/gaia/}, which is set to provide unparalleled astrometric precision, allowing us to remove unresolved binaries and obtain much more precise absolute magnitudes for these objects.

Building on other studies, we have shown that an optical - infrared color is an excellent metallicity indicator. If a similar relation can be inferred for metallicity and distance in the Sloan ($g'$,$r'$,$i'$,$z'$) standard bandpasses, this would be a resource for future surveys like LSST, which is expected to uncover billions of M dwarfs. Towards this end, we have recently begun an effort to gather the necessary data on a subset of the stars presented here to explore this question. We plan to publish and release these data in a future publication.

\acknowledgments
We thank the anonymous referee for his or her insight and suggestions in reviewing this manuscript. E.R.N. was supported throughout this work by a National Science Foundation Graduate Research Fellowship and A.W.M. by the Harlan J. Smith Fellowship from the University of Texas at Austin. The MEarth Team gratefully acknowledges funding from the David and Lucille Packard Fellowship for Science and Engineering (awarded to D.C.). This material is based upon work supported by the National Science Foundation under grants AST-0807690, AST-1109468, and AST-1004488 (Alan T. Waterman Award). This publication was made possible through the support of a grant from the John Templeton Foundation. The opinions expressed in this publication are those of the authors and do not necessarily reflect the views of the John Templeton Foundation. This research has made extensive use of data products from the Two Micron All Sky Survey, which is a joint project of the University of Massachusetts and the Infrared Processing and Analysis Center/California Institute of Technology, funded by NASA and the NSF, NASAs Astrophysics Data System (ADS), and the SIMBAD database, operated at CDS, Strasbourg, France. The MEarth team would also like to acknowledge Eric Mamajek who initially suggested for us to engage in this work. The authors of this paper would also like to thank Zachory Berta-Thompson, whose comments, suggestions, and extensive work with the MEarth observatory and data are invaluable.

\bibliography{bibliography}

\clearpage

\begin{table}
\begin{center}
\caption{Quantum Efficiency (QE) of the MEarth CCD, and Filter Throughput (Table Stub)}
\label{filter_table}
\begin{tabular}{crrrr} 
\tableline\tableline
Wavelength (nm) & QE (CCD) & Transmission (MEarth Filter) & Product \\
\hline
650 & 0.907 & 0.000 & 0.000 \\
660 & 0.901 & 0.000 & 0.000 \\
670 & 0.895 & 0.000 & 0.000 \\
680 & 0.887 & 0.000 & 0.000 \\
690 & 0.879 & 0.005 & 0.005 \\
700 & 0.869 & 0.059 & 0.051 \\
710 & 0.857 & 0.311 & 0.267 \\
720 & 0.845 & 0.666 & 0.563 \\
730 & 0.831 & 0.866 & 0.720 \\
740 & 0.816 & 0.941 & 0.768 \\
... & ... & ... & ... \\
\tableline
\tableline 
\tableline
\end{tabular}
\end{center}
\end{table}

\begin{table}
\begin{center}
\caption{Nightly Photometric Coefficients (Equation \ref{equation1}) (Table Stub) }
\label{coefficient_table}
\begin{tabular}{crrrr} 
\tableline\tableline
Date & Color Coefficient & Airmass Coefficient & Zero Point & Telescope \\
\tableline
2011-10-13 & 0.0006 & 0.062 & 20.695 & tel03 \\
2011-10-13 & 0.0033 & 0.058 & 20.448 & tel01 \\
2011-10-14 & -0.0040 & 0.066 & 20.512 & tel04 \\
2011-10-14 & 0.0002 & 0.068 & 20.699 & tel07 \\
2011-10-14 & 0.0050 & 0.048 & 20.515 & tel05 \\
2011-10-15 & -0.0062 & 0.086 & 20.700 & tel03 \\
2011-10-15 & -0.0049 & 0.081 & 20.744 & tel08 \\
2011-10-15 & -0.0028 & 0.078 & 20.760 & tel06 \\
2011-10-15 & -0.0013 & 0.075 & 20.430 & tel01 \\
2011-10-15 & -0.0007 & 0.086 & 20.699 & tel02 \\
... & ... & ... & ... & .. \\
\tableline 
\tableline
\end{tabular}
\end{center}
\end{table}

\newpage

\begin{table}
\begin{center}
\caption{$MEarth$ Magnitudes of Landolt Standards (Table Stub)}
\label{standard_table}
\begin{tabular}{crrr} 
\tableline\tableline
Landolt Star & $m_{MEarth}$ & err \\
\tableline
SA92 263 & 10.703 & 0.033 \\
SA92 348 & 11.443 & 0.032 \\
SA92 409 & 9.264 & 0.036 \\
SA92 342 & 11.096 & 0.031 \\
SA92 335 & 11.809 & 0.032 \\
SA92 253 & 12.762 & 0.035 \\
SA92 250 & 12.363 & 0.035 \\
SA92 245 & 11.982 & 0.037 \\
SA95 43 & 10.179 & 0.034 \\
SA95 41 & 12.866 & 0.036 \\
... & ... & ... \\
\tableline 
\tableline
\end{tabular}
\end{center}
\end{table}

\newpage

\begin{sidewaystable}
\begin{center}
\caption{Observed Properties of the MEarth Targets (Table Stub)}
\label{big_table}
\begin{tabular}{crrrrrrrrrr} 
\tableline\tableline
2MASS ID & LSPM-N ID & $m_{MEarth}$ & err & $m_J$ & err & $m_H$ & err & $m_K$ & err  \\ 
\hline
00011579+0659355 & J0001+0659 & 13.038 & 0.013 & 11.286 & 0.022 & 10.741 & 0.028 & 10.418 & 0.021 \\
00020623+0115360 & J0002+0115 & 14.516 & 0.029 & 12.168 & 0.022 & 11.543 & 0.023 & 11.129 & 0.023 \\
00074264+6022543 & J0007+6022 & 10.335 & 0.012 & 8.911 & 0.021 & 8.332 & 0.024 & 8.046 & 0.02 \\ 
00085391+2050252 & J0008+2050 & 10.391 & 0.012 & 8.87 & 0.027 & 8.264 & 0.031 & 8.01 & 0.024 \\
00085512+4918561 & J0008+4918 & 12.636 & 0.017 & 10.864 & 0.022 & 10.32 & 0.021 & 9.98 & 0.018 \\
00113182+5908400 & J0011+5908 & 11.727 & 0.015 & 9.945 & 0.023 & 9.393 & 0.026 & 9.093 & 0.021 \\
00115302+2259047 & J0011+2259 & 10.267 & 0.011 & 8.862 & 0.021 & 8.308 & 0.036 & 7.987 & 0.021 \\ 
00125716+5059173 & J0012+5059 & 13.177 & 0.021 & 11.406 & 0.022 & 10.818 & 0.027 & 10.517 & 0.022 \\ 
00131578+6919372 & J0013+6919 & 9.858 & 0.013 & 8.556 & 0.024 & 7.984 & 0.02 & 7.746 & 0.02 \\
00133880+8039569 & J0013+8039W & 8.971 & 0.031 & 7.756 & 0.034 & 7.131 & 0.047 & 6.904 & 0.02 \\ 
... & ... & ... & ... & ... & ... & ... & ... & ... & ... \\

\tableline 
\tableline
\end{tabular}

\end{center}
\end{sidewaystable}

\begin{sidewaystable}
Table 4 stub (cont)
\begin{center}
\begin{tabular}{crrrrrrrr} 
\tableline\tableline
$\pi$ (\arcsec) &  err & Source & Spectral Type & Source &[Fe/H] (spec) & err & [Fe/H] (photo) \\
\hline
 0.057 & 0.0026 & \citet{2014ApJ...784..156D} & -- & -- & -0.083 & 0.12 & -0.209 \\
 0.059 & 0.0047 & \citet{2014ApJ...784..156D} & M9 & \citet{2011AJ....141...97W} & -- & -- & 0.012 \\ 
 0.069 & 0.002 & \citet{2014ApJ...784..156D} & M3.8 & \citet{2009ApJ...699..649S} & -- & -- & 0.053 \\
 0.054 & 0.00135 & \citet{2014AJ....147...85R} & M5 & \citet{2014MNRAS.443.2561G} & 0.12 & 0.12 & 0.300 \\
 -- & -- & -- & M5.5 & \citet{2003AJ....126.3007R} & -- & -- & -- \\
 0.108 & 0.00052 & \citet{2009AJ....137..402G} & M5.5 & \citet{2015AA...577A.128A} & -- & -- & -0.232 \\
 0.060 & 0.0018 & \citet{2014ApJ...784..156D} & M3.5 & \citet{2009ApJ...699..649S} & -- & -- & 0.140 \\
0.038 & 0.0031 & \citet{2014ApJ...784..156D} & M6.4 & \citet{2009ApJ...699..649S} & -- & -- & 0.112 \\ 
0.053 & 0.0029 & \citet{2014ApJ...784..156D} & M 3 & \citet{1995AJ....110.1838R} & -- & -- & -0.046 \\ 
0.051 & 0.00175 & \citet{2007AA...474..653V} & M1.5 & \citet{2015AA...577A.128A} & -- & -- & -0.032 \\ 
... & ... & ... & ... & ... & ... & ... & ... \\
\tableline 
\tableline
\end{tabular}
 Trigonometric Parallaxes are taken from \citet{2000AJ....120.1106B, 2007ApJ...668L.155M, 1992AJ....103..638M, 2009AJ....137.4109L, 2013A&A...551A..48A, 2007AA...474..653V, 2014ApJ...784..156D, 2002AJ....124.1170D, 2008AJ....136..452G, 1993AJ....105.1571H, 1999AJ....118.1086B, 2014AJ....147...85R, 2009AJ....137..402G, 2010AJ....140..897R, 2012ApJ...758...56S, 1995gcts.book.....V, 2003AJ....125..354R, 2010AstL...36..576K, 2006AJ....132.2360H, 2010A&A...514A..84S, 2007A&A...464..787S}
 
 Spectral types are taken from \citet{2003AJ....125.1598L, 2000MNRAS.311..385G, 2000AJ....120..447K, 2004AJ....128..463R, 2002AJ....123.2828C, 1991ApJS...77..417K, 1997AJ....113..806G, 2011AJ....141...97W, 2006AJ....132..866R, Landolt09, 2002AJ....123.3434L, 2007AJ....133..439C, 2000AJ....120.1085G, 2007AJ....133.2825R, 1991AJ....101..662B, 2008AJ....136.1290R, 2010AJ....140..897R, 1997PASP..109..849G, 2002AJ....123.2002H, 2005AJ....130.1871B, 2014MNRAS.443.2561G, 1994AJ....108.1437H, 2003AJ....126.3007R, 2013AJ....145...52M, 2005A&A...433..151Z, 2003AJ....126.2421C, 2010ApJ...716.1522S, 1995AJ....110.1838R, 1995AJ....109..797K, 2004AJ....128.2460H, 2006AJ....132.2360H, 2002ApJ...581L..47L, 2013AJ....145..102L, 1997ApJ...476..311K, 2002AJ....124..519R, 1996A&AS..116..467M, 2009ApJ...699..649S, 2005PASP..117..676R, 2005A&A...442..211S,2015AA...577A.128A}
 
Blend Flag 3 are taken from \citet{2000AJ....120.1106B, 2013AJ....146..156D, 2007ApJ...654..558D, 1960PASP...72..125W, 2014ApJ...789..102J, 1998AJ....115.2053G, 2015ApJS..216....7B, 1984IAUC.3989....1H, 2003ApJ...587..407C, 2001AJ....121.3259M, 2011ApJ...742..123I, 1997AJ....114..388H, 1997ApJ...475..604S, 2005A&A...435L...5F, 1936ApJ....84R.359K, 2001AJ....121..489R, 2004A&A...425..997B, 2013A&A...549A.109B, 1976PUSNO..24c...1D, Hipparcos_original, 2010ApJ...716.1522S, 2006A&A...460L..19M, 2011AJ....141..166H, 2004ApJ...617.1323P, 2008ApJ...678..463I, 2006MNRAS.368.1917L, 1999ApJ...512..864H, 1996AJ....111..365G, 1999A&A...344..897D, 2008MNRAS.384..150L, 2012ApJ...754...44J, 2001AJ....122.3466M, 2009ApJ...701.1436I, 2005ASPC..338..288P, 2009ApJ...704..975J, 2006ApJ...649..389P} as well as IRTF, MEarth, and SDSS images.
\end{center}
\end{sidewaystable}

\begin{sidewaystable}
Table 4 stub (cont)
\begin{center}
\begin{tabular}{crrrrrrrrr} 
\tableline\tableline
Distance Modulus (m-M) & Blend Flag 1 & Blend Flag 2 & Blend Flag 3 & Source  \\
\hline
-- & 0 & 0 & 0 & -- \\
-- & 0 & 0 & 0 & -- \\
-- & 1 & 0 & 1 & \citet{2014ApJ...789..102J} \\
-- & 0 & 0 & 1 & \citet{2004AA...425..997B} \\
0.921 & 0 & 0 & 0 & -- \\
-- & 1 & 0 & 0 & -- \\
 -- & 0 & 0 & 0 & -- \\
 -- & 0 & 0 & 0 & -- \\
 -- & 0 & 0 & 1 & \citet{2001AJ....122.3466M} \\
 -- & 0 & 0 & 0 & -- \\
 ... & ... & ... & ... & ...  \\
\tableline 
\tableline
\end{tabular}
\end{center}
\end{sidewaystable}

\begin{sidewaystable}
Table 4 stub (cont)
\begin{center}
\begin{tabular}{crrrr} 
\tableline\tableline
 First datapoint (JD) & Last datapoint (JD) & N data points & Over-luminous? \\
\hline
2455858.794 & 2456838.905  & 24 & 0 \\
2455867.658 & 2456822.961 & 23 & 0 \\
2455848.728 & 2456702.623 & 517 & 0 \\
2455850.702 & 2456836.954 & 91 & 0 \\
2455857.730 & 2456830.888 & 27 & 0 \\
2455874.621 & 2456834.960 & 105 & 0 \\
2455850.741 & 2456836.886 & 417 & 0 \\
2455851.675 & 2456836.958 & 76 & 0 \\
2455855.821 & 2456834.922 & 278 & 0 \\
2455857.770 & 2456839.899 & 36 & 0 \\
 ... & ... & ... & ... \\
\tableline 
\tableline
\end{tabular}
\end{center}
\end{sidewaystable}

\newpage

\begin{table}
\begin{center}
\caption{Color-Magnitude Sequence at [Fe/H] = -0.1 and [Fe/H] = 0.0 (Table Stub)}
\label{main_sequence_track}
\begin{tabular}{crrrr} 
\tableline\tableline
$m_{MEarth} - m_K$  & $M_{K} ([Fe/H] = -0.1)$ & $M_{K} ([Fe/H] = 0.0)$ \\
\tableline
2.100 & 6.452 & 5.992 \\
2.103 & 6.487 & 6.028 \\
2.105 & 6.487 & 6.063 \\
2.108 & 6.523 & 6.098 \\
2.110 & 6.558 & 6.134 \\
2.113 & 6.558 & 6.169 \\
2.116 & 6.593 & 6.205 \\
2.118 & 6.629 & 6.240 \\
2.123 & 6.664 & 6.275 \\
2.126 & 6.664 & 6.311 \\
... & ... & ... \\
\tableline 
\tableline
\end{tabular}
\end{center}
\end{table}

\newpage

\end{document}